\documentclass[prl,aps,groupedaddress]{revtex4}
\usepackage{amssymb,amsmath,amsfonts,epsfig,graphicx,latexsym,bm,color}

\begin{document}

\title{Dynamical phase transition in the open Dicke model}
\author{J. Klinder, H. Ke{\ss}ler, M. Wolke, L. Mathey, and A. Hemmerich \footnote{e-mail: hemmerich@physnet.uni-hamburg.de} }
\affiliation{Institut f\"{u}r Laser-Physik, Universit\"{a}t Hamburg, Luruper Chaussee 149, 22761 Hamburg, Germany}

\maketitle
\noindent
\textbf{The Dicke model with a weak dissipation channel is realized by coupling a Bose-Einstein condensate to an optical cavity with ultra-narrow bandwidth. We explore the dynamical critical properties of the Hepp-Lieb-Dicke phase transition by performing quenches across the phase boundary. We observe hysteresis in the transition between a homogeneous phase and a self-organized collective phase with an enclosed loop area showing power law scaling with respect to the quench time, which suggests an interpretation within a general framework introduced by Kibble and Zurek. The observed hysteretic dynamics is well reproduced by numerically solving the mean field equation derived from a generalized Dicke Hamiltonian. Our work promotes the understanding of nonequilibrium physics in open many-body systems with infinite range interactions.}
\hfill\break \hfill\break
\noindent 
dynamical phase transition $\mid$ critical behavior $\mid$ Dicke model $\mid$ quantum gas $\mid$ cavity QED
 \hfill\break \hfill\break

\noindent
While equilibrium phases in quantum many-body systems have been explored for a long time with great success, non-equilibrium phenomena in such systems are far less well understood \cite{Pol:11}. A paradigm for exploring non-equilibrium dynamics is the quench scenario, where a system parameter is subjected to a sudden change between two values associated with different equilibrium phases. Quantum degenerate atomic gases with their unique degree of control are particularly adapted for experimental quench studies \cite{Lew:07, Blo:08}. For isolated quantum many-body systems a wealth of theoretical and experimental investigations of quench dynamics has appeared recently \cite{Cal:06, Kol:07, Sil:08, Hey:13, Sad:06, Kin:06, Gri:12, Che:12}. A natural extension of such studies is to consider driven open systems, where dynamical equilibrium states can arise via a competition between dissipation and driving, and non-equilibrium transitions between such phases can occur as a function of some external control parameter \cite{Hoh:77, Die:10, Tor:10, Sie:13}. A nearly ideal experimental platform for this endeavor are quantum degenerate atomic gases subjected to optical high finesse cavities, where the usual extensive control in cold gas systems can be combined with a precisely engineered coupling to the external bath of vacuum radiation modes \cite{Rit:13}.

\begin{figure}
\includegraphics[scale=0.4, angle=0, origin=c]{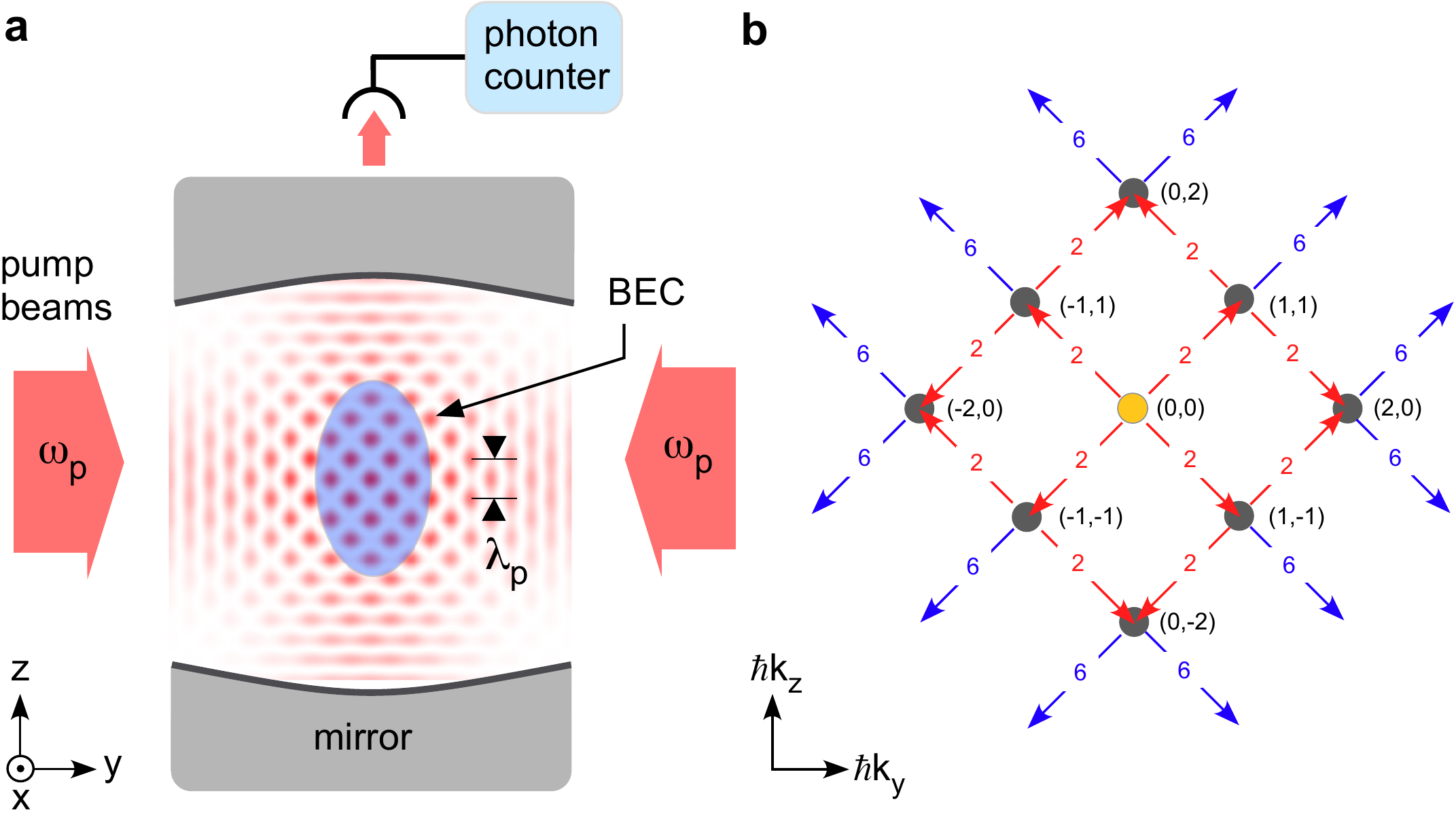}
\caption{\label{Fig.1} (a) Experimental scheme. The intra-cavity intensity is indicated by the red pattern with the antinodes corresponding to the locations, where the atoms are localized. The frequency and wavelength of the pump beams are denoted by $\omega_p$ and $\lambda_p$, respectively. (b) The available atomic momentum states that can couple to the condensate are indicated by their momentum components (n,m) in the $y$- and $z$-directions in units of the pump photon momentum $\hbar k$. Arrows of the same color identify scattering processes involving the same kinetic energy transfer, denoted in units of the recoil energy $E_{\mathrm{rec}} \equiv \hbar \omega_{\mathrm{rec}} \equiv \hbar^2 k^2/ 2m$ ($m =$ atomic mass)  by the numbers on top of the arrows.}
\label{fig:setup}
\end{figure}

Here, we study a dynamical phase transition in the open Dicke model emulated in an atom-cavity system prepared near zero temperature. The Dicke model is a paradigmatic scenario of quantum many-body physics, still subject to intensive research despite a more than half a century long history \cite{Dic:54, Dic:64, Hep:73, Gil:78, Bow:79, Gro:82, Ema:03, Nag:10, Bha:12, Bas:12, Pia:13, Kul:13}. It describes the interaction of $N$ two-level atoms with a common mode of the electromagnetic radiation field. Hepp and Lieb have pointed out already in the seventies that, upon varying the coupling strength, this model possesses a second order equilibrium quantum phase transition between a homogeneous phase, in which each atom interacts separately with the radiation mode, and a collective phase in which all atomic dipoles align to form a macroscopic dipole moment \cite{Hep:73, Gro:82}. It has been early suspected that the critical properties of the externally pumped open Dicke model should give rise to non-linear hysteretic behavior in dynamical experiments \cite{Gil:78, Bow:79}. The dynamical properties of the open Dicke model and of related many-body atom-cavity systems in presence of dissipation are subject of extensive recent theoretical research \cite{Gop:09, Nag:10, Stra:11, Bha:12, Tor:13, Kul:13}.  

With the atomic levels chosen to be momentum states of the external motion, the open Dicke model has been recently implemented experimentally by coupling a Bose-Einstein condensate (BEC) to a high finesse resonator pumped by an external optical standing wave \cite{Bau:10}. A transition from a \textit{homogeneous} phase (consisting of the condensate with no photons in the cavity) into a \textit{collective} phase (with the atoms forming a density grating trapped in a stationary intra-cavity optical standing wave) was observed at a critical pump strength close to the expected equilibrium transition boundary. A related transition with thermal atoms has been studied in earlier work \cite{Do:02, Bla:03}. In the formation of the \textit{collective} phase the $Z_2$ symmetry, associated with two possible grating configurations shifted with respect to each other by half an optical wavelength, is spontaneously broken \cite{Bla:03, Bau:11}. In Ref.~\cite{Bau:10} the cavity dissipation rate was more than two orders of magnitude larger than the single photon recoil frequency with the consequences that the intra-cavity light field adiabatically adjusts to the evolution of the atomic distribution on a microsecond time scale.

\begin{figure}
\includegraphics[scale=0.5, angle=0, origin=c]{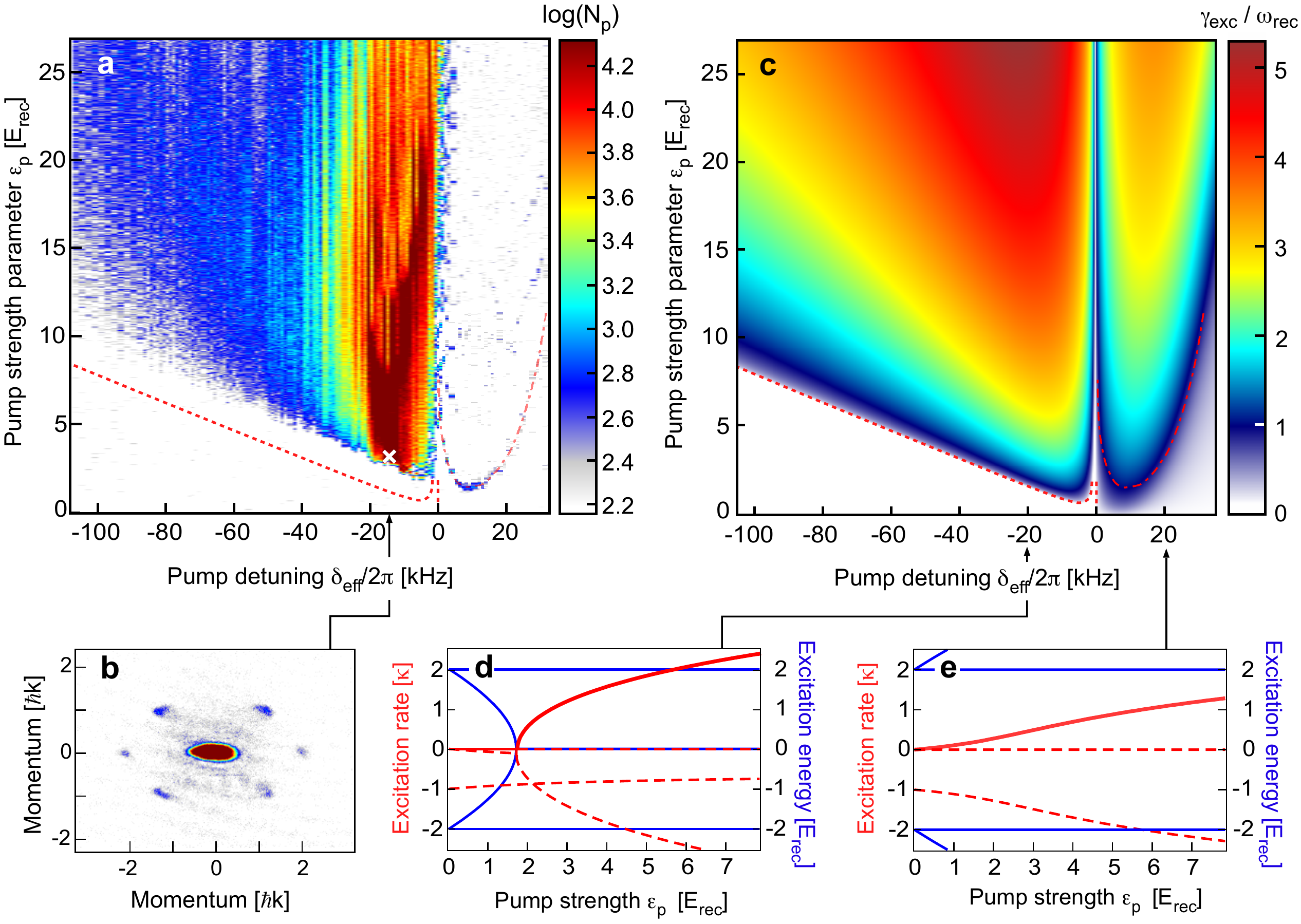}
\caption{\label{Fig.2} (a) The observed intra-cavity photon number $N_p$ is plotted versus the pump strength parameter $\varepsilon_p$ and the pump detuning $\delta_{\mathrm{eff}}$. (b) A momentum spectrum is shown recorded at the point in (a) marked by the white cross. The dashed red line indicates the equilibrium Dicke phase boundary obtained from the stability analysis illustrated in (c). (c) Plot of the maximal excitation rate $\gamma_{\mathrm{exc}}$ calculated from a stability analysis for the homogeneous phase of the Dicke model. For negative $\delta_{\mathrm{eff}}$, the equilibrium Dicke phase transition line is highlighted by the red dashed line. For positive $\delta_{\mathrm{eff}}$, the dashed dotted red line indicates the contour $\gamma_{\mathrm{exc}} \approx 0.8\, \omega_{\mathrm{rec}}$, where superradiant pulses are observed in (a). In (d) and (e) for $\delta_{\mathrm{eff}} = \pm 2 \pi \times 20\,$~kHz the real (solid blue lines) and imaginary (dashed red lines) parts of all eigenvalues of the stability matrix are plotted. The maximal values of the latter are highlighted by the solid red lines.}
\label{fig:setup}
\end{figure}

In the present work, the main innovation is the use of a cavity with ultra-narrow bandwidth on the order of the single photon recoil frequency \cite{Wol:12, Kes:14}. The time scales for dissipation of the intra-cavity field and the coherent atomic evolution are similar. We can thus dynamically access the non-adiabatic regime, where both quantities are not in equilibrium and hence explore non-equilibrium critical properties of the Dicke model in quench experiments. For different signs of the effective detuning $\delta_{\mathrm{eff}}$ of the pump field with respect to the cavity resonance, we observe fundamentally different behavior. Remarkably, the Hepp-Lieb-Dicke transition, observed for negative detuning $\delta_{\mathrm{eff}}<0$, shows a dynamical hysteresis. The resulting hysteresis loop encloses an area, which exhibits power law dependence upon the duration of the quench across the phase boundary and maintains non-zero values even at quench time scales by far slower than the dynamical time scales of the underlying single particle Hamiltonian. We interpret this finding in the framework of the Kibble-Zurek model \cite{Kib:76, Zur:85, Cam:14}. Our observations are consistent with solutions of the mean-field equations associated with a Dicke Hamiltonian \cite{Dic:54}. A second interesting consequence is the observation of an instability boundary for positive detuning $\delta_{\mathrm{eff}}>0$. The physics of this instability, which is also predicted by the Dicke model but not much discussed in the literature, resembles cavity-assisted matter wave superradiance \cite{Ino:99}, recently observed to prevail at any value of $\delta_{\mathrm{eff}}$ for single-sided pumping \cite{Kes2:14}. The system is excited by a cascade of successive superradiant pulses to form coherent superpositions of discrete momentum states with the zero momentum condensate mode practically depleted. No stationary intra-cavity field is formed in this case. At the boundary between the two regimes within a narrow interval around zero detuning $\delta_{\mathrm{eff}}\approx0$ we find that the atoms cannot scatter photons at all even at large values of the pump strength.

\textbf{Experimental scheme}. In our experiment, outlined in Fig.~1(a), a cigar-shaped BEC of $\mathrm{^{87}Rb}$-atoms is prepared such that its long axis is well aligned with the axis of a longitudinal mode of a high finesse ($\mathcal{F} = 3.44 \pm 0.05 \times 10^5$) optical standing wave resonator. The atoms are exposed to an optical standing wave oriented perpendicularly with respect to the cavity axis. The strength of this external pump wave is parametrized by the depth $\varepsilon_p$ of the associated light shift potential in units of the recoil energy, which is determined spectroscopically (for details see \textit{SI Appendix}). The frequency $\omega_{p}$ of the pump wave is far detuned from the relevant atomic resonances, such that the interaction with the atoms is dispersive with negligible spontaneous emission (for details see \textit{SI Appendix}). The cavity possesses an extremely low dissipation rate associated with the loss of photons. The field decay rate $\kappa = 2\pi \times 4.45\pm 0.05\,$kHz is smaller than twice the recoil frequency $2\,\omega_{\mathrm{rec}} = 2\pi \times 7.1\,$kHz, which corresponds to the kinetic energy transferred to a resting atom by scattering a pump photon into the cavity and hence sets the time scale of the atomic motion. As a consequence, the choice of $\omega_{p}$ relative to the resonance frequency $\omega_{c}$ of the empty cavity selects only a small fraction of the atomic momentum states, illustrated in Fig.~1(b), to be resonantly coupled. For a uniform atomic sample and left circularly polarized light, the TEM$_{00}$ resonance frequency is dispersively shifted by an amount $\delta_{-} = \frac{1}{2} N_a \, \Delta_{-}$ with an experimentally determined light shift per photon $\Delta_{-} \approx \,2\pi \times 0.36 \pm 0.04\,$Hz. With $N_a = 10^5$ atoms $\delta_{-} = 2\pi\times 18$~kHz, which amounts to $4\,\kappa$, i.e., the cavity operates well within the regime of strong cooperative coupling (for details see \textit{SI Appendix}).

\begin{figure}
\includegraphics[scale=0.5, angle=0, origin=c]{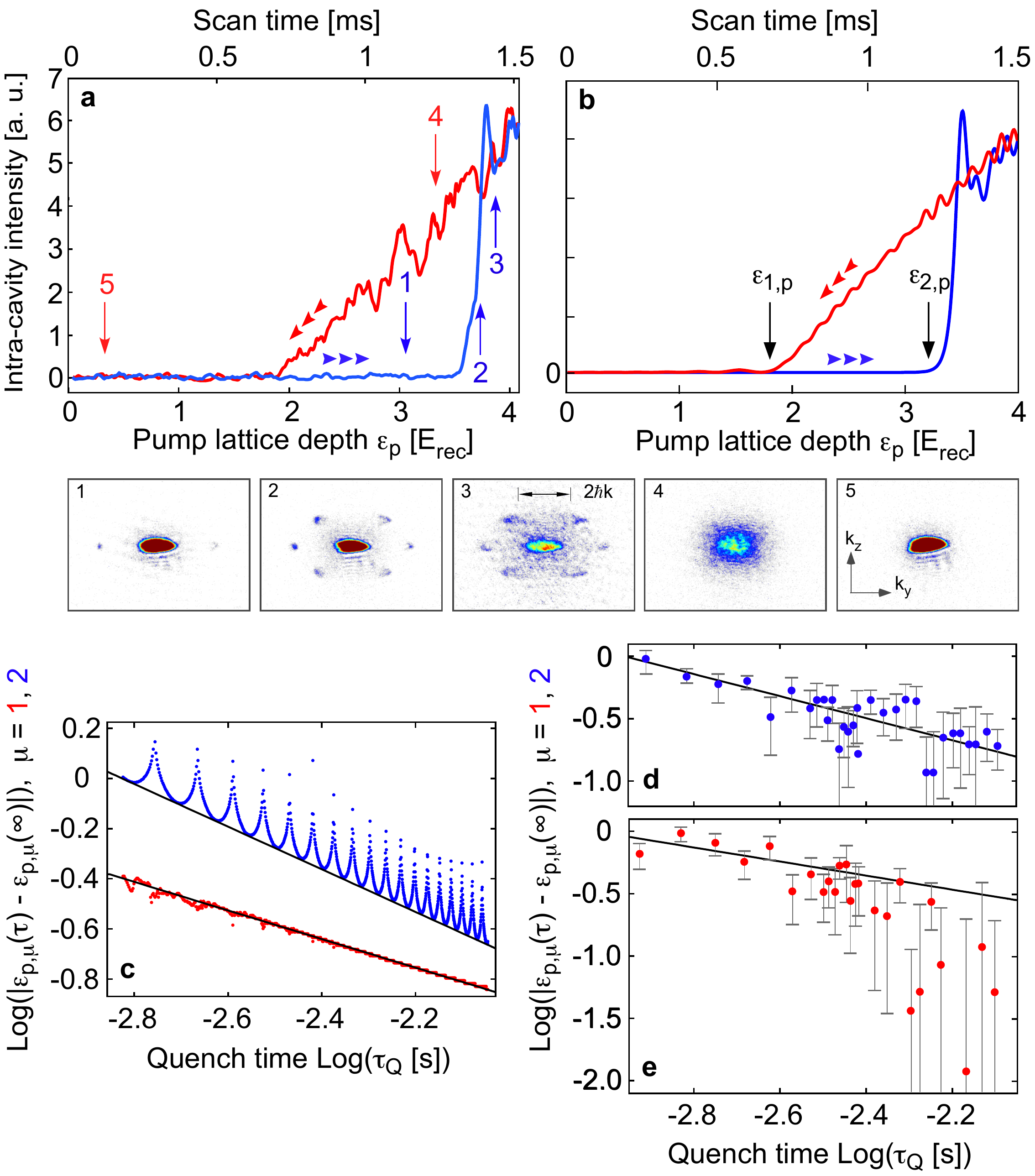}
\caption{\label{Fig.3} (a) For fixed $\delta_{\mathrm{eff}} = - 2 \pi \times 17.5\,$~kHz the intra-cavity intensity is plotted with the pump strength ramped from 0 to $4\,E_{\mathrm{rec}}$ in 1.5 ms (blue line) and back (red line). Below (a), a series of consecutively numbered momentum spectra is shown (1-5), recorded at increasing times during the $\varepsilon_p$-ramp, indicated by the correspondingly numbered arrows in (a). (b) A mean field calculation according to (a) for a homogeneous, infinite system without collisional interaction. In (c) mean field calculations of $\varepsilon_{p,1}$ (lower graph, red dots) and $\varepsilon_{p,2}$ (upper graph, blue dots) are shown. The solid lines show power laws with exponents $n_1=-0.57$ and $n_2=-0.85$ in the lower and upper graph, respectively. The measured dependence of the critical values $\varepsilon_{p,\mu}, \mu \in \{1,2\}$ upon the quench time $\tau_Q$ is shown for $\mu=2$ in (d) and for $\mu=1$ in (e). The solid lines repeat the power laws found in the mean field calculations in (c). The error bars reflect the standard deviations for 10 measurements.}
\label{fig:setup}
\end{figure}

\textbf{Hepp Lieb Dicke transition}. For negative detuning $\delta_{\mathrm{eff}} \equiv \omega_{p}-\omega_{c} -\delta_{-}<0$, above a critical value of $\varepsilon_p$, a stationary intra-cavity light field builds up and the atoms are captured in the ground state of the light shift potential formed by the interference of the intra-cavity field and the pump wave. In Fig.~2(a), the observed intra-cavity power is plotted versus $\delta_{\mathrm{eff}}$ and $\varepsilon_p$. This graph is obtained by linearly ramping up $\varepsilon_p$ at a rate $1.4 \,E_{\mathrm{rec}}\,$ms$^{-1}$ at fixed values of $\delta_{\mathrm{eff}}$. The formation of the optical lattice is readily seen by detecting Bragg maxima in the atomic momentum spectra obtained by a time-of-flight method. In Fig.~2(b) this is shown for the position in the phase diagram in (a) marked by the white cross. Within a narrow channel around $\delta_{\mathrm{eff}}=0$, scattering of photons is entirely suppressed. Above $\varepsilon_p \approx 5$ this channel becomes so narrow that our limited accuracy of the pump frequency (about $\pm\,200$~Hz) does not provide sufficient resolution. This may be understood as follows: For $|\delta_{\mathrm{eff}}|$ exceeding $\kappa$, the intra-cavity field is driven in phase with the pump field. Hence, interference of the two fields yields a square lattice potential with minima arranged on a Bravais lattice spanned by the primitive vectors $(\hat{y}\pm\hat{z})\,\lambda_p /2$ with $\hat{y},\hat{z}$ denoting the unit vectors in $y$- and $z$-directions. The density grating formed by trapping atoms in these minima (corresponding to the intensity maxima in Fig.~1(a)) satisfies the Bragg condition for scattering photons from the pump field into the cavity. When $|\delta_{\mathrm{eff}}|$ becomes smaller than $\kappa$, the relative phase between the intra-cavity field and the pump field approaches $\pi/2$ for $|\delta_{\mathrm{eff}}| \rightarrow 0$. This suppresses the interference between both fields. As a result, the unit cell develops a second minimum, which approaches equal depth for $\delta_{\mathrm{eff}}\rightarrow 0$. The associated density grating, now populating both classes of minima, no longer supports Bragg scattering of pump photons into the cavity, and hence the intra-cavity field and the density grating collapse. 

The basic structure of the observations in Fig.~2(a) can be understood as follows. At low pump powers the dynamics of the system may be described by the Heisenberg equations for the matter and light variables associated with the Dicke Hamiltonian with an additional term describing dissipation of the cavity light field at a rate $\kappa$ (for details see \textit{SI Appendix}). These equations possess a stationary solution describing the homogeneous phase, when all atoms populate the condensate mode at zero intra-cavity intensity. Linearization around this solution yields a stability matrix, whose eigenvalues are readily calculated. Their real parts denote the excitation spectrum while their imaginary parts denote the corresponding exponential excitation rates. Hence, if one of the eigenvalues attains a positive imaginary part, the homogeneous phase becomes unstable. The maximum of the imaginary parts of all eigenvalues, denoted by $\gamma_{\mathrm{exc}}$, is plotted in Fig.~2(c) versus $\delta_{\mathrm{eff}}$ and $\varepsilon_p$. Fig.~2(d) and (e) show the excitation frequencies (blue solid lines) and the corresponding excitation rates (red dashed lines) along vertical sections in (c) with fixed detunings $\delta_{\mathrm{eff}} = \pm 2 \pi \times 20\,$~kHz. The solid red lines highlight the maxima $\gamma_{\mathrm{exc}}$ of the excitation rates. For negative detuning, below a critical value of $\varepsilon_p$ the homogeneous phase is predicted to be stable. As the critical value $\varepsilon_{p,c}$ is approached, a softening of one of the excitation modes is seen in Fig.~2(d) (descending solid blue line starting at $2\,E_{\mathrm{rec}}$) and as $\varepsilon_{p,c}$ is passed, $\gamma_{\mathrm{exc}}$ attains positive values (solid red line), indicating instability of the homogeneous phase. In Fig.~2(c), $\varepsilon_{p,c}(\delta_{\mathrm{eff}})$ is highlighted by a red dashed line, which indicates the expected equilibrium Dicke phase transition boundary. This boundary is also registered in the data in Fig.~2(a). 

\begin{figure}
\includegraphics[scale=0.5, angle=0, origin=c]{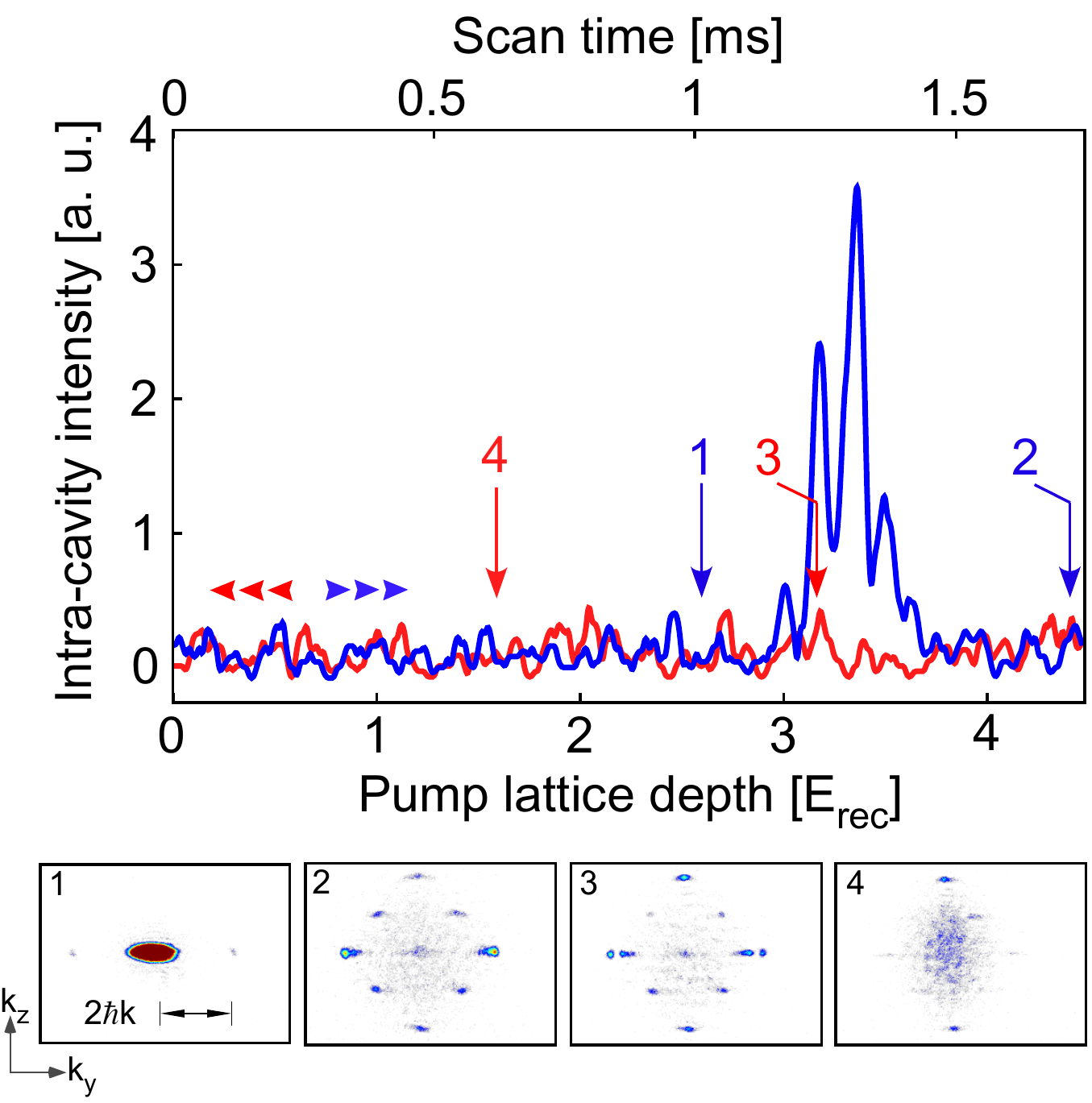}
\caption{\label{Fig.4} The intra-cavity intensity is plotted for fixed detuning $\delta_{\mathrm{eff}} = 2 \pi \times 4\,$~kHz with the pump lattice depth $\varepsilon_p$ ramped from 0 to $4.5\,E_{\mathrm{rec}}$ in 1.7~ms (blue trace) and back (red trace). At the lower edge of the graph, four numbered momentum spectra are shown, taken at consecutive times, indicated by the numbered arrows. }
\label{fig:setup}
\end{figure}

\textbf{Observation of hysteresis}. As is seen in Fig.~2(a), the phase transition for increasing $\varepsilon_p$ is observed at values of $\varepsilon_p$ beyond the equilibrium phase boundary. As also noted in Ref.~\cite{Bha:12} this should be expected because sufficiently large values of $\gamma_{\mathrm{exc}}$ must be reached in Fig.~2(c) before the system can leave the homogeneous phase in a given time. A more complete picture is provided in Fig.~3, where the transition through the phase boundary is studied for negative detuning $\delta_{\mathrm{eff}} = - 2 \pi \times 17.5\,$~kHz in more detail. In Fig.~3(a) $\varepsilon_p$ is ramped up from 0 to 4 in 1.5 ms and back to 0 again in 1.5 ms. The solid blue and red lines show the observed intra-cavity intensity for the increasing and decreasing sections of the ramp, respectively. Note that this quantity measures the depth of the intra-cavity lattice emerging in the collective phase and hence corresponds to the square of the order parameter for the Dicke phase transition. A significant hysteresis is observed. For increasing $\varepsilon_p$ a sudden jump of the intra-cavity intensity arises on a time scale corresponding to the cavity decay rate. On the way back, the intra-cavity intensity is smoothly tuned to zero. In the center row of the figure (below (a) and (b)), a series of consecutively numbered momentum spectra is shown, recorded at different instances of time during the $\varepsilon_p$-ramp, indicated by the correspondingly numbered arrows in (a). As the intra-cavity intensity assumes finite values, a coherent optical lattice is formed (arrow 2), as is seen from the occurrence of higher order Bragg peaks. As the lattice depth grows (arrows 3 and 4), tunneling amplitudes decrease, and the relatively increased collisional interaction acts to reduce particle number fluctuations resulting in a loss of coherence. When ramping back to small values of $\varepsilon_p$, the BEC is recovered with no notable atom loss and only few low energy Bogoliubov excitations (arrow 5). 

In Fig.~3(b) a mean field calculation (based upon a Dicke Hamiltonian, see \textit{SI Appendix}) is shown for a homogeneous, infinite system without collisional interaction, which shows the same signatures as observed in (a) including dynamical details as the oscillation of the red trace around $\varepsilon_p \approx 2.5$ and the overshooting of the blue trace around $\varepsilon_p \approx 3.5$. The observed hysteresis appears fundamentally different from that known to occur in conventional bistable systems, where discontinuities arise for both critical values, where the system becomes unstable. We do not find a discontinuity at the lower critical value $\varepsilon_{p,1}$ in Figs.~3(a,b), however, the system always follows the blue curve, when this point is passed with increasing $\varepsilon_p$, irrespective of the duration $\tau_Q$ of the applied $\varepsilon_p$-ramp. For increasing cavity bandwidths our mean field calculations predict that the area enclosed by the hysteresis decreases and finally is obscured by increasing optomechanical oscillations at the phase boundary (see also Fig.~12(a) in Ref.~\cite{Bha:12}).

\textbf{Power law scaling}. The dependence of the threshold values $\varepsilon_{p,1}$ and $\varepsilon_{p,2}$ for the dynamical transitions in Fig.~3(a) and (b) upon the quench time $\tau_Q$ is studied in (c), (d) and (e). These quantities are determined as those values of $\varepsilon_p$, where the intra-cavity intensity assumes five percent of its maximal value reached for $\varepsilon_p = 4$. In (c) the values of $\Delta \varepsilon_{p,\mu}(\tau_Q)\equiv \varepsilon_{p,\mu}(\tau_Q) - \varepsilon_{p,\mu}(\tau_Q=\infty)$ ($\mu \in \{1,2\}$), calculated from curves as that shown in (b), are plotted versus $\tau_Q$. As shown by the solid lines, the $\tau_Q$-dependences follow power laws $\Delta \varepsilon_{p,{\mu}}(\tau_Q) \propto {\tau_Q}^{n_{\mu}}$ with $n_1=-0.57$ and $n_2=-0.85$. The phase offset of the sharp resonances occurring periodically at a frequency $\Omega = 0.682\, \omega_{\mathrm{rec}}$ in the upper blue trace depends on the specific choice of a small initial excitation, necessary to drive the system out of the homogeneous phase, which is provided by quantum and thermal fluctuations in the experiment. The exponents $n_1, n_2$ turn out independent of the exact initial conditions (for details see \textit{SI Appendix}). In (d) and (e) we plot the experimentally observed values of $\Delta \varepsilon_{p,\mu}(\tau_Q)$ with $\mu=2$ and $\mu=1$, respectively. The solid lines repeat the power laws found in the calculations in (c) with $n_2=-0.85$ in (d) and $n_1=-0.57$ in (e). While in (d) the data nicely agree with the power law behavior, in (e) this is only the case for the first half of the plot. At later times the data points assume an exponential rather than a power law decay, which is in accordance with the observation that for long ramp times at the end of the descending ramp notable particle loss sets in. Our observations of power law behavior of $\Delta \varepsilon_{p,\mu}(\tau_Q)$ suggests an interpretation within the universal model introduced by Kibble and Zurek \cite{Kib:76, Zur:85, Cam:14}, which applies for second order phase transitions in isolated many-body systems. According to this model a quench between two phases is approximated by a succession of an adiabatic approach towards and a departure from the equilibrium critical point $\varepsilon_{p,c}$ conjoined by a diabatic passage through the critical point, where the dynamics is completely frozen. Furthermore, a power law dependence for the relaxation time is assumed, i.e., $\tau(\varepsilon_{p}) \propto |\varepsilon_{p}-\varepsilon_{p,c}|^{-z_{\mu} \nu_{\mu}}$ with $\mu \in \{1,2\}$ if $\varepsilon_{p}<\varepsilon_{p,c}$ and $\varepsilon_{p}>\varepsilon_{p,c}$, respectively. The identification of $\Delta \varepsilon_{p,\mu}$ with the lower and upper bounds of the diabatic region around $\varepsilon_{p,c}$ then leads to the prediction that $z_{\mu} \nu_{\mu}= -(1 + \frac{1}{n_{\mu}})$, i.e., in our system: $z_{1} \nu_{1} = 0.75$, $z_{2} \nu_{2} = 0.18$ (for details see \textit{SI Appendix}). A deeper understanding of these values would require a comprehensive extension of the concept of universality to the case of driven open systems \cite{Sie:13}.

\textbf{Matter wave superradiance}. 
In the $\delta_{\mathrm{eff}}>0$ region of Fig.~2(a), matter wave superradiance prevails \cite{Ino:99, Kes2:14}. Short superradiant pulses with a duration on the order of the intra-cavity photon life time are emitted by the cavity, if $\varepsilon_p$ reaches a critical instability boundary, highlighted by a red dashed dotted line in Fig.~2(a). The atoms, initially populating the condensate mode at zero momentum, are thereby scattered into superpositions of higher momentum states. This excitation is irreversible and cannot be removed by ramping $\varepsilon_p$ back to zero. As Fig.~2(c) and (e) show, $\gamma_{\mathrm{exc}}$ always exceeds zero for non-zero $\varepsilon_p$. Hence, the homogeneous phase is everywhere unstable. The observed instability boundary corresponds to a contour of constant $\gamma_{\mathrm{exc}}\approx 0.8\,\omega_{\mathrm{rec}}$ (highlighted by the dashed dotted red line in Fig.~2(c), replotted from (a)). The value of this constant increases with increasing speed of the applied $\varepsilon_p$-ramp. Our mean-field calculations show that the increase of $\gamma_{\mathrm{exc}}$ with $\varepsilon_p$ significantly reduces for increasing cavity bandwidth. Hence, in the experimental scenario of Ref.~\cite{Bau:10}, matter wave superradiance is expected to occur only at observation times or values of $\varepsilon_p$ much larger than realized there.

Fig.~4 shows the intra-cavity intensity for fixed positive detuning $\delta_{\mathrm{eff}} = 2 \pi \times 4\,$~kHz with $\varepsilon_p$ ramped from 0 to $4.5\,E_{\mathrm{rec}}$ in 1.7~ms and back to 0 again in 1.7~ms. At the lower edge of the graph, four numbered momentum spectra are shown, taken at consecutive times, indicated by the numbered arrows. At the chosen value of $\delta_{\mathrm{eff}}$ the scattering processes indicated by the red arrows in Fig.~1(b), associated with $2\,E_{\mathrm{rec}}$ energy transfer, are nearly resonantly driven, while the processes indicated by blue arrows in Fig.~1(b) are significantly detuned, and hence do not contribute. Accordingly, at the threshold value $\varepsilon_p \approx 3\,E_{\mathrm{rec}}$ a short superradiant light pulse is emitted from the cavity, after which a large fraction of the atoms is transferred to the $(\pm 2, 0) \hbar k$ momentum states (arrow (2)). These states do not support a matter wave Bragg grating necessary to maintain further scattering. Hence, the cavity field falls to zero and the momentum states propagate towards the trap edges, which are reached in about 1.25~ms corresponding to the trap frequency of 200~Hz in the $y$-direction. Reflection of the higher momentum components at the anharmonic trap edges and collisions yield a rapid broadening of the momentum distribution (arrow (4)).

\textbf{Conclusions}. We have studied quench dynamics in the open Dicke model emulated by strongly coupling an atomic Bose-Einstein condensate to an optical cavity providing an extremely narrow bandwidth. Our experiment exhibits a uniquely controlled paradigm of non-equilibrium many-body dynamics in presence of dissipation, which appears ideal for quantitative confrontations with theory also beyond mean field approximations. We hope that this work will stimulate new theoretical efforts to better understand the connection between non-linear dynamics and statistical mechanics in open many-body systems.
\\ \\
\textbf{ACKNOWLEDGMENTS}. This work was partially supported by DFG-SFB 925 and DFG-GrK1355. We are grateful to Michael Thorwart, Reza Bakhtiari, Duncan O'Dell, and Helmut Ritsch for useful discussions.

\section{Supplementary Information}

\textbf{Parameters of Bose-Einstein condensate}.
A cigar-shaped Bose-Einstein condensate (BEC) with Thomas-Fermi radii $(3.1, 3.3, 26.8)\,\mu$m and $N_a \approx 10^5$ $\mathrm{^{87}Rb}$-atoms, prepared in the upper hyperfine component of the ground state $|F=2,m_F=2\rangle$, is confined by three centimeter-sized solenoids \cite{Han:06, Kli:10} arranged in a quadrupole Ioffe configuration \cite{Ess:98}, thus providing a magnetic trap with a nonzero bias field parallel to the $z$-axis with trap frequencies $\omega / 2\pi = (215.6 \times 202.2 \times 25.2)\,$Hz. The particle number in the atomic sample is measured by absorption imaging and by recording the cavity resonance shift due to forward scattering of a probe beam, coupled through one of the cavity mirrors. We thus find less than $10\,\%$ shot to shot fluctuations.
\\ \\
\textbf{Cavity parameters}.
The high finesse of the standing wave cavity ($\mathcal{F} = 3.44 \pm 0.05 \times 10^5$) together with the narrow beam waist ($w_0 \approx 31.2 \pm 0.1\, \mu$m) yield a Purcell factor $\eta_\mathrm{c} \equiv \frac{24\,\mathcal{F}}{\pi\,k^2 w_0^2} \approx 44\pm 0.7$ ($k \equiv 2\pi/\lambda$, and $\lambda=\,$ wavelength of the pump light) \cite{Pur:46, Tan:11}. Due to the mirror separation of $48.93\pm0.002\,$mm, the cavity exhibits an extremely low bandwidth of $\kappa = 2\pi \times 4.45\pm 0.05\,$kHz, which is smaller than $2\,\omega_\mathrm{rec}$, with $\omega_\mathrm{rec} = \hbar k^2/2 m = 2\pi \times 3.55\,$kHz denoting the recoil frequency. The cavity is oriented parallelly to the $z$-axis, such that the BEC is well matched to the mode volume of its TEM$_{00}$-modes with its elongated axis aligned parallel to the cavity axis. Note that the BEC extends across approximately 130 lattice sites of the intra-cavity standing wave and thus position fluctuations in the BEC preparation process yield only small population fluctuations between adjacent sites. For a uniform atomic sample the resonance frequency for right ($+$) and left ($-$) circular photons is shifted due to the dispersion of a single atom by an amount $\Delta_{\pm} / 2$ with $\Delta_{\pm} = \frac{1}{2} \eta_\mathrm{c} \kappa \,\Gamma \left( \frac{f_{1,\pm}}{\delta_1} + \frac{f_{2,\pm}}{\delta_2}\, \right)$ and $\delta_{1,2}$ denoting the pump frequency detunings with respect to the relevant atomic D$_{1,2}$ lines at $795.0\,$nm and $780.2\,$nm \cite{Tan:11}. $\Gamma = 2 \pi \times 6$~MHz is the intra-cavity field decay rate and the decay rate of the $5\mathrm{P}$ state of $\mathrm{^{87}Rb}$, respectively. The prefactors $f_{1,\pm}$ and $f_{2,\pm}$ account for the effective line strengths of the D$_{1}$- and D$_{2}$-line components connecting to the $|F=2, m_F=2 \rangle$ ground state. The values of these factors are $(f_{1,-}, f_{2,-}) = (\frac{2}{3},\frac{1}{3})$ and $(f_{1,+}, f_{2,+}) = (0, 1)$. 

The quoted expressions for $\Delta_{\pm}$ use the rotating wave approximation and assume that the contributions from different transition components may be added. Finite size effects of the atomic sample and deviations of the intra-cavity field geometry from a plane wave are neglected. A more realistic value, used in our work, is obtained experimentally: The dispersive resonance shift for $N_a$ atoms and left polarized light $\delta_{-} = \frac{1}{2} N_a \Delta_{-}$ is measured by coupling a weak left polarized probe beam through one of the cavity mirrors to the TEM$_{00}$-mode. Its frequency is tuned across the resonance with and without atoms. At sufficiently low power levels of the probe the resonance is not affected by spatial structuring of the atoms due to back-action of the cavity field and hence merely results from the dispersion of the homogeneous sample. Accounting for the particle number $N_a$, known from absorption imaging with about $10\,\%$ precision, we find $\Delta_{-} \approx - 2\pi \times 0.36 \pm 0.04\,$Hz and $\Delta_{+} \approx - 2\pi \times 0.16\pm 0.02$~Hz. Hence with $N_a = 10^5$ atoms $\delta_{-} = 2\pi\times 18$~kHz, which amounts to $4\,\kappa$, i.e., the cavity operates well within the regime of strong cooperative coupling. 
\\ \\
\textbf{Pump lattice parameters}. The pump lattice with $w_p = 80 \pm 0.4\,\mu$m radius is oriented along the $y$-axis, i.e., perpendicularly with respect to its weakly confined $z$-axis. Its linear polarization is oriented parallelly to the $x$-axis and it operates at a wavelength $\lambda = 803\,$nm, i.e., with $8\,$nm detuning to the red side of the D$_{1}$-transition of $\mathrm{^{87}Rb}$. The pump strength is specified in terms of the magnitude of the antinode light shift $\varepsilon_p \geq 0$ induced by the pump lattice in units of the recoil energy $E_\mathrm{rec} = \hbar \omega_\mathrm{rec}$. In order to calibrate the pump strength, the BEC is adiabatically loaded into the pump lattice and the excitation spectrum is recorded and compared to a numerical band calculation. This yields a relative (absolute) uncertainty of $\varepsilon_p$ of $1\%$ (10\%).

Our experiments require to tune the pump frequency with sub-kilohertz resolution across the resonance frequency of the TEM$_{00}$-mode interacting with the BEC. This is accomplished as follows (see also Ref.~\cite{Kes:14}): A reference laser operating at 803 nm is locked on resonance with a TEM$_{11}$-mode, which provides a cloverleaf-shaped  transverse profile. This mode exhibits a nodal line at the cavity axis such that the interaction with the BEC, which is positioned well in the center of the TEM$_{00}$-mode, is suppressed with respect to the TEM$_{00}$-mode by a geometrical factor $9 \times 10^{-5}$. Adjusting right circular polarization for the reference beam and hence $\sigma^+$-coupling of the BEC yields another suppression factor $\approx 0.43$. The pump laser, matched to couple the TEM$_{00}$-mode, is locked with an offset frequency of about $2.5\,$GHz to the reference laser. This offset is tunable over several MHz such that the vicinity of the resonance frequency of the TEM$_{00}$-mode can be accessed. 
\\ \\
\textbf{Detection of cavity photons}.
The light leaking out of the cavity is split into orthogonal circular polarization components and the photons of each component are counted with $56 \%$ quantum efficiency. The right circular photons, predominantly belong to the TEM$_{11}$-mode used to operate the stabilization of the pump beam frequency with respect to the cavity resonance (for details see Ref.~\cite{Kes:14}). Only a small fraction of these photons arises in the TEM$_{00}$-mode and results from the scattering of pump photons. In our experiments the ratio between left and right circularly polarized photons found in the TEM$_{00}$-mode was about 4. The photon counting signal is binned within a time-window of $4 \mu$s and a variable number of data sets is averaged. 
\\ \\
\textbf{Mean field model}.
We consider a BEC of two-level atoms scattering light from an external standing wave mode with the scalar electric field amplitude $\alpha_{p}(t) \cos(ky)$ (pump mode) into a cavity mode with the scalar electric field $\alpha(t) \cos(kz)$. Neglecting collisional interaction the system is described by the set of mean field equations \cite{Rit:13}
\begin{eqnarray}
\label{eq:mean_field_model}
i\,\frac{\partial}{\partial t} \psi(y,z,t) &=& \left(-\frac{\hbar}{2m}\left[\frac{\partial^2}{\partial y^2}+\frac{\partial^2}{\partial z^2}\right] +  \Delta_0 |\alpha(t) \cos(kz)+\alpha_{p}(t) \cos(ky)|^2 \right) \psi(y,z,t) 
\\ \nonumber
i\,\frac{\partial}{\partial t} \alpha(t) &=& \left(-\delta_c + \Delta_0 \langle \cos^2(kz)\rangle_{\psi} - i \kappa\right) \alpha(t) + \Delta_0 \langle \cos(kz)  \cos(ky)\rangle_{\psi} \,\alpha_p(t) \, ,
\end{eqnarray}
with the matter wave function $\psi$ normalized to $N_a$ particles, and the electric fields normalized such that $|\alpha_{p}|^2$ and $|\alpha|^2$ denote the number of photons in the pump mode and the cavity mode, respectively. The light shift per intra-cavity photon is denoted by $\Delta_0$. $\langle \dots \rangle_{\psi}$ indicates integration over the BEC volume weighted with $|\psi|^2$. A plane wave expansion of $\psi(y,z,t)$ with respect to the relevant $(y,z)$-plane yields the corresponding scaled momentum space equations
\begin{eqnarray}
\label{eq:mean_field_equations}
\nonumber
i\,\frac{\partial}{\partial t} \phi_{n,m} &=& \omega_{\mathrm{rec}}\, \left(n^2+m^2 - \frac{1}{2} |\beta|^2\, - \frac{1}{2} \epsilon_p \right) \phi_{n,m}
\\ \nonumber
&-&\,\frac{1}{4} \omega_{\mathrm{rec}}\,|\beta|^2 \left(\phi_{n,m-2}+\phi_{n,m+2}\right)- \frac{1}{4} \omega_{\mathrm{rec}}\,\epsilon_p \left(\phi_{n-2,m}+\phi_{n+2,m}\right)
\\ \nonumber
&+&\,\frac{1}{2} \omega_{\mathrm{rec}} \, \sqrt{\epsilon_p} \, \mathrm{Im}(\beta) \left(\phi_{n-1,m-1}+\phi_{n+1,m-1}+\phi_{n-1,m+1}+\phi_{n+1,m+1}\right)
\\  \nonumber
\\
i\,\frac{\partial}{\partial t} \beta &=& \left[-\delta_{\mathrm{eff}} + \frac{1}{2} N_a \Delta_0 \sum_{n,m} \mathrm{Re}[\phi_{n,m}\phi_{n,m+2}^{*}]  - i \kappa \right] \, \beta
\\ \nonumber
&-&i\,\frac{1}{8} N_a \Delta_0 \,\sqrt{\epsilon_p} \, \sum_{n,m} \phi_{n,m}(\phi_{n+1,m+1}^{*}+\phi_{n+1,m-1}^{*})+\phi_{n,m}^{*}(\phi_{n+1,m+1}+\phi_{n+1,m-1}) \, ,
\end{eqnarray}
with $ \phi_{n,m}$ denoting the normalized ($\sum_{n,m}|\phi_{n,m}|^2 = 1$) amplitude of the momentum state $(n,m)\, \hbar k$. Upon the assumption of negative $\Delta_0$ the intra-cavity field $\beta$ is scaled such that $|\beta|^2 = - |\alpha|^2 \Delta_0 / \omega_{\mathrm{rec}}$ denotes the magnitude of the induced anti-node light-shift in units of the recoil energy. The pump strength parameter $\epsilon_p \equiv - |\alpha_p|^2 \Delta_0 / \omega_{\mathrm{rec}}$ is defined as the antinode light-shift induced by the pump wave in units of the recoil energy. The effective detuning is $\delta_{\mathrm{eff}}  \equiv \delta_c - \frac{1}{2} N_a \Delta_0$ with the detuning $\delta_c$ between the pump frequency and the empty cavity resonance. Eq.~(\ref{eq:mean_field_equations}) is the mean-field approximation to the Heisenberg equation for a Dicke Hamiltonian generalized to the case of a collection of $N_a$ identical multi-level systems each consisting of the momentum states $\phi_{n,m}$. The additional term $i\kappa$ accounts for damping of the intra-cavity light field. The conventional two-level Dicke-Hamiltonian \cite{Dic:54} arises, if only the two most relevant matter modes $\phi_{0,0}$ and $\phi \equiv \frac{1}{2}\left(\phi_{1,1} + \phi_{1,-1}+\phi_{-1,1} + \phi_{-1,-1} \right)$ are accounted for. In our recoil selective cavity set-up, this approximation is well justified since initially the atoms populate the BEC mode $\phi_{0,0}$ and near resonant coupling via the cavity is practically limited to $\phi$, which corresponds to the motional state excited, if a single photon from the standing pump wave is scattered into the cavity. Note that within the sub-space spanned by the states $\phi_{\pm 1,\pm 1}$ the superposition $\phi$ represents the minimal energy state, because the associated density grating localizes the particles in the minima of the intra-cavity lattice potential induced by photon scattering into the cavity.

A steady state solution of Eqs.~(\ref{eq:mean_field_equations}) is the homogeneous phase $\beta = 0$ and $\phi_{n,m} = \delta_{n,0} \,\delta_{m,0}$, which describes the unperturbed condensate with no photons in the cavity. The stability properties of this solution may be studied by reducing Eqs.~(\ref{eq:mean_field_equations}) to the two matter modes $\phi_{0,0}$ and $\phi$. Switching to a basis such that the condensate has zero energy and neglecting its depletion, i.e., $\phi_{0,0} \approx 1$, one finds the system of linear equations
\begin{eqnarray}
\label{eq:4ModeModel}
i\,\frac{\partial}{\partial t} \left(\begin{array}{c} \beta \\ \beta^{*} \\ \phi \\ \phi^{*} \end{array} \right)
= 
\left(\begin{array}{cccccc}
- \delta_{\mathrm{eff}} - i \kappa & 0 & i \, \lambda_{1}  & i \, \lambda_{1}  \\
  0 & \delta_{\mathrm{eff}} - i \kappa & i \, \lambda_{1} &  i \, \lambda_{1} \\ 
  -i \lambda_{2} & i \lambda_{2} & 2\omega_{\mathrm{rec}} & 0 \\
    i \lambda_{2} & -i \lambda_{2} & 0 & -2\omega_{\mathrm{rec}}  \\
\end{array} \right) 
\left(\begin{array}{c} \beta \\ \beta^{*} \\ \phi \\ \phi^{*} \end{array} \right)
\end{eqnarray}
with the coupling parameters $\lambda_{1} \equiv -\frac{1}{4} \,N_a \Delta_0 \, \sqrt{\epsilon_p}$ and $\lambda_{2} \equiv \,\omega_{\mathrm{rec}}\,\sqrt{\epsilon_p}$, which formally resembles a Schr{\"o}dinger equation for a four-level system with a non-Hermitian Hamiltonian \cite{Ben:13}. It is equivalent to the mean field approximation of the Heisenberg equation obtained from the Dicke Hamiltonian after applying the Holstein-Primakoff transformation, introducing the thermodynamic limit, and adding cavity dissipation \cite{Ema:03}. If the imaginary part of one of the eigenvalues of the matrix on the right hand side of Eq.~(\ref{eq:4ModeModel}) is positive, an exponential instability arises and hence the system is rapidly driven away from the homogeneous phase. For negative detuning $\delta_{\mathrm{eff}}<0$ the instability boundary is the known equilibrium Dicke phase boundary.

In order to compare experimental observations with the model in Eq.~(\ref{eq:mean_field_equations}) and Eq.~(\ref{eq:4ModeModel}), the experimental parameters $\Delta_{\pm}, \varepsilon_p$ and the model parameters $\Delta_{0}, \epsilon_p$ must be connected accounting for the fact that in the model two-level atoms are assumed and the vectorial character of the electric field is neglected. In the experiment, the strongest coupling to the atoms arises for left circular light with respect to the natural quantization axis fixed by the magnetic offset field along the $z$-axis. Hence, we identify $\Delta_{0} = \Delta_{-}$. Inside the cavity, the linear $\hat{x}$-polarization of the pump beam may be decomposed into equally strong left and right circular components with respect to the $z$-axis. Only the left circular component can scatter into the left circularly polarized cavity mode. Hence, the light shift $\varepsilon_p$ induced by the pump beam in the experiment is related to the number of pump photons $|\alpha_p|^2$ used in the model description by $\varepsilon_p = - |\alpha_p|^2 (\Delta_{+} + \Delta_{-}) / \omega_{\mathrm{rec}}$ and thus $\varepsilon_p/\epsilon_p = (\Delta_{+} + \Delta_{-}) /\Delta_{0} = 1.44$. A more involved description, which is deferred to forthcoming work, should account for two orthogonal polarization modes of the cavity operating with different effective detunings. 

In Fig.~2(c) of the main text the maximum of the imaginary parts of the four eigenvalues of the matrix on the right hand side of Eq.~(\ref{eq:4ModeModel}) is plotted versus $\delta_{\mathrm{eff}}$ and $\varepsilon_p$. For $\delta_{\mathrm{eff}} = \pm 2 \pi \times 20$~kHz the real and imaginary parts of all eigenvalues are plotted in Fig.~2(d) and (e), respectively. Figures 3(b) and (c) in the main text were obtained by solving Eqs.~(\ref{eq:mean_field_equations}) including all modes with $-4 \leq n,m \leq 4$. A small initial deviation from $\phi_{0,0}(0) =1$ is required in order to leave the unstable homogeneous phase. In the experiment, this deviation is naturally provided by thermal or quantum fluctuations. We assumed that the first excited modes $(\pm 1,\pm 1)\,\hbar k$ are populated according to a Boltzman factor with a temperature $T = 0.2 \,T_c$ ($T_c =$ critical temperature of the BEC). Hence, we set $\phi_{0,0}(0) = \cos(\theta)$, $\phi_{\pm 1, \pm 1}(0) = e^{i \xi} \sin(\theta)/2$ with $\theta = \arctan(2\,e^{-\hbar \omega_{\mathrm{rec}} / k_B T})$, and $\phi_{n,m}(0) = 0$ if $|n|,|m| >1$. The choice of equal $\phi_{\pm 1, \pm 1}(0)$ corresponds to low energy excitations for which the atoms are localized in the intra-cavity light shift potential with a spatial phase determined by the positions of the cavity mirrors and the position of the pump wave. The choice of $\xi$ determines the phase of the oscillations in the upper trace of Fig.~3(c), which are washed out, if an average over $\xi$ is applied. For $T<T_c$, the exponents of the power law behavior in this figure does not show notable dependence on $\xi$ or the value of $T$. Note that the higher orders $\phi_{n,m}$ with $|n|,|m|>1$ remain small in the calculation of Fig.~3(b) and the hysteresis is also reproduced in the simplified case $-1 \leq n,m \leq 1$, which corresponds to a description in terms of the Dicke model for a collection of two-level systems. 
\\ \\
\textbf{Power law scaling, relation to Kibble Zurek model}.
We consider a quench across the equilibrium Dicke phase transition implemented by tuning the pump strength parameter $\varepsilon_{p}(t) = \varepsilon_{p,c} + (-1)^{\mu} \frac{\Delta \varepsilon}{\tau_{Q}}\,(t-t_c)$ across the critical value $\varepsilon_{p}(t_c)=\varepsilon_{p,c}$, with $\Delta \varepsilon$ denoting the interval of $\varepsilon_{p}(t)$ scanned during the quench time $\tau_Q$. For $\mu \in \{1,2\}$ we identify $\varepsilon_{p,c} = \varepsilon_{p,\mu}(\infty)$, where $\varepsilon_{p,\mu}(\infty)$ denote the threshold values found in a quench with negative ($\mu =1$) or positive ($\mu =2$) slope in the limit of infinite $\tau_Q$. According to our experimental observations, the quantities $\Delta \varepsilon_{p,\mu}(\tau_Q)\equiv \varepsilon_{p,\mu}(\tau_Q) - \varepsilon_{p,\mu}(\infty)$ follow power laws, i.e., $\Delta \varepsilon_{p,\mu} \propto \tau_Q^{n_{\mu}}$ with $n_1= -0.57$ and $n_2= -0.85$. At this point we argue in the spirit of the Kibble Zurek model \cite{Cam:14}, that the time lag between the threshold value for the transition to occur in a quench of duration $\tau_Q$ and the equilibrium critical point, i.e. $\tau_Q\,\Delta \varepsilon_{p,\mu}(\tau_Q)$, equals the relaxation time $\tau$ of the system, and hence $\tau \propto \tau_Q^{n_{\mu}+1}$. As a second input from the Kibble Zurek scenario, we assume a power law scaling $\tau \propto \Delta \varepsilon_{p,\mu}^{-z_{\mu} \nu_{\mu}}$. This results in the relation $z_{\mu} \nu_{\mu} = -(1+1/n_{\mu})$, and hence $z_{1} \nu_{1}= 0.75$ and $z_{2} \nu_{2}= 0.18$.

\end{document}